\documentstyle[aps,twocolumn]{revtex}  
 
\input psfig
 
\begin{document} 
  
\draft 
  
\title{Electric field effects in Fibonacci superlattices} 
 
\author{Mario Castro$^{\dag}$ and Francisco 
Dom\'{\i}nguez-Adame$^{\ddag}$} 
 
\address{Departamento de F\'{\i}sica de Materiales, 
Facultad de F\'{\i}sicas, Universidad Complutense, 
E-28040 Madrid, Spain \\ and \\ Grupo Interdisciplinar de 
Sistemas Complicados, Escuela Polit\'ecnica Superior, 
Universidad Carlos III, E-28911 Legan\'{e}s, Madrid, Spain \\
\bigskip}

\author{\small\parbox{14cm}{\small
 We present a throughout study of transmission and localization
properties of Fibonacci superlattices, both in flat band conditions
and subject to homogeneous electric fields perpendicular to the
layers. We use the transfer matrix formalism to determine the
transmission coefficient and the degree of localization 
of the electronic states. We find that the fragmentation pattern
of the electronic spectrum is strongly modified when the electric
field is switched on, this effect being more noticeable as the
system length increases. We relate those phenomena to  
field-induced localization of carriers in Fibonacci superlattices.
\\[3pt]PACS numbers: 
71.50.$+$t, 
72.15.Rn,   
73.20.Dx}   
}\address{}\maketitle

\section{Introduction} 
 
The fabrication of aperiodic se\-mi\-con\-duc\-tor su\-per\-lattices 
(SLs), arranged according to the Fibonacci \cite{Merlin1} and Thue-Morse 
\cite{Merlin2} sequences, has given rise to a growing interest in their 
exotic electronic and transport properties 
\cite{Merlin3,Laruelle,Yamaguchi,Toet,Hirose,Munzar,%
Macia1,Adame1,Adame2,Adame3}.  Theoretical studies demonstrate that 
ideal aperiodic SLs should exhibit a highly-fragmented and fractal-like 
electronic spectrum \cite{Laruelle,Macia1,Adame1,Adame2}.  This 
self-similar spectrum is observable even when unintentional 
imperfections arising during the growth process are considered 
\cite{Adame3}.  The electronic states associated to this peculiar 
spectrum are no longer Bloch states and also present fractal-like 
properties \cite{Kohmoto,Macia2}, although they extend over the entire 
sample, what is most important for subsequent applications in actual 
devices, these novel properties have been experimentally observed.  For 
instance, photoluminescence excitation spectroscopy at low temperature 
reveals the existence of a fragmented density of states consistent with 
theoretical predictions \cite{Yamaguchi}. 
 
Most devices work under bias conditions and, consequently, a complete 
characterization of electronic states in aperiodic SLs subject to an 
applied electric field is indeed needed.  In this way, it has 
been recently demonstrated that time-dependent coherent oscillations of 
electronic wavepackets induced by the homogeneous electric field (Bloch 
oscillations) are absent in these SLs \cite{Diez1}.  This is to be 
compared with periodic SLs, where Bloch oscillations have been predicted 
and detected in Ga$_{1-x}$Al$_{x}$As \cite{Leo,Feldmann}.  In this paper 
we address the study of the energy spectrum and transmission property of 
Fibonacci SLs (FSLs) subject to an applied electric field. We aim to 
get deep insight into the interaction of fractal-like electronic
states of FLSs with the external field, which clearly is more complex
than in the case of periodic SLs. We will 
focus our attention on the effects of this electric field on the 
fragmented spectrum.  Moreover, we will also discuss the competition 
between the long-range order of the FSL and the localization effects of 
the external electric field.  To be specific, we consider the scattering 
problem of an electron impinging on a quasi-one dimensional FSL. 
Transport properties for different electron energies will be described 
by means of the transmission coefficient since this magnitude is 
directly related to the conductance of the sample.  To get an estimation 
of the degree of localization of the electronic state as a function of 
energy, we will use the inverse participation ratio (IPR) to be defined 
below. These two magnitudes will be shown below to be enough for our
present purposes. 
 
\section{Model} 
 
We consider quantum-well based SLs with the same barrier thickness $b$ 
in the whole sample.  The height of the {\em n\/}th barrier with center 
at $z_{n}=na$ barrier is given by the conduction-band offset, $z$ being 
the coordinate in 
the growth direction and $a>b$ the separation between neighbouring 
barriers.  We will focus on electronic states close to the bandgap with 
${\bf k}_{\perp}=0$ and neglect nonparabolicity effects hereafter, so 
that the Ben Daniel-Duke Hamiltonian suffices to describe those states. 
For the sake of simplicity, we consider FSLs with narrow barriers, 
namely we assume strong coupling between quantum wells.  From a 
mathematical point of view we require that $b\to 0$ whereas the area of 
the barrier remain unchanged ($\delta$-function limit).  Therefore, the 
envelope-functions for electronic states satisfy the following 
time-independent Schr\"{o}dinger equation 
\begin{equation} 
\left[ -\,{\hbar^2\over 2m}\,{d^2\phantom{x}\over dz^2} + 
\sum_{n=1}^{N} \> V_{n}a\delta(z-na) - eFz \right]\,\psi(z) = 
E\,\psi(z), 
\label{1} 
\end{equation} 
where $V_{n}a$ is the strength of the $\delta$-function, $N$ is the 
number of barriers 
and $F$ is the electric field.  We take the origin of electron energies 
at the conduction-band edge in the quantum-wells. {}FSLs can be grown 
starting from two different barriers with strengths $V$ and $V'$, 
arranged according to the Fibonacci sequence.  The Fibonacci sequence 
$S_n$ is generated by appending the $n-2$ sequence to the $n-1$ one, 
i.e., $S_n={S_{n-1}S_{n-2}}$, where $S_{0}=V'$ and $S_{1}=V$.  Thus, 
finite and self-similar quasiperiodic SLs are obtained by $n$ successive 
applications of these rules containing $N=F_n$ barriers arranged 
according $V\,V'\,V\,V\,V'\ldots$ The Fibonacci numbers are generated 
from the recurrence law $F_n=F_{n-1}+F_{n-2}$, starting with 
$F_0=F_1=1$. 
 
To proceed, we use the transfer-matrix method to calculate the 
transmission coefficient, in a similar fashion to the case of periodic 
SLs \cite{Sun}.  We define the length $\ell = \ell(F) \equiv 
(\hbar^2/2meF)^{1/3}$, the dimensionless parameter and $\lambda = 
\lambda(F,E) \equiv (2m/\hbar^2e^2F^2)^{1/3}E$, and the 
dimensionless variable $y = y(z,F,E) \equiv (2/3) \left[ \lambda + 
z/\ell \right]^{3/2}$, in order to obtain the solution of (\ref{1}) in 
terms of the Hankel functions of first and second kind 
\begin{equation} 
\psi_n(z)=A_n y^{1/3} H^{(1)}_{1/3}(y) + B_n y^{1/3} H^{(2)}_{1/3}(y), 
\quad x_{n-1}<x<x_{n}, 
\label{2} 
\end{equation} 
where the constants $A_n$ and $B_n$ are to be determined from the boundary 
conditions. The coefficients in the free-field regions ($x < x_0$ and $x> x_N$)
are related through the transfer matrix $T_N$ as follows:
\begin{equation}
\left( \begin{array}{c} A_N \\ B_N \end{array} \right) =
T_N \left( \begin{array}{c} A_0 \\ B_0 \end{array} \right) \equiv
\prod_{n=N}^{1} M_n 
\left( \begin{array}{c} A_0 \\ B_0 \end{array} \right),
\label{3}
\end{equation}
where 
\begin{equation}
M_n = \left( \begin{array}{cc} 
1+\alpha_n f_n/i_n & \alpha_n g_n/i_n \\
 -\alpha_n h_n/i_n & 1-\alpha_n f_n/i_n 
\end{array} \right)
\label{4}
\end{equation}
is the site transfer matrix and for brevity we have defined 
$\alpha_n=(2m\ell V_na/\hbar^2)(3y_n/2)^{-1/3}$ with $y_n=y(na,F,E)$, 
$f_n=H^{(2)}_{1/3}(y)H^{(1)}_{-2/3}(y)-H^{(1)}_{1/3}(y)H^{(2)}_{-2/3}(y)$,
$g_n=H^{(2)}_{1/3}(y)H^{(1)}_{1/3}(y)$, $h_n=[H^{(2)}_{1/3}(y)]^{(2)}$ 
and $i_n=[H^{(1)}_{1/3}(y)]^{(2)}$. The total transfer matrix ${\cal T}_N$
relates the amplitudes of plane waves in the free-field regions and the
transmission coefficient is then determined from the relationship
$\tau=|\mbox{det}({\cal T}_N)/({\cal T}_N)_{22}|^{2}$. 

As mentioned in the Introduction, we use the IPR to determine the
degree of localization of the wave function for different incoming
energies. The amplitude distribution of the electronic states can 
be characterized by the moments associated to the measure defined
in the system by us (in our case the probability of finding the 
electron at a given point of the lattice). We then use the second 
moment of this distribution, which is nothing but the so-called
IPR, defined as follows
\begin{equation}
\mbox{IPR} = \frac{ \sum_{n=1}^{N}\> |\psi_n(na)|^{4} }
{\left( \sum_{n=1}^{N}\> |\psi_n(na)|^{2} \right)^{2} }.
\label{5}
\end{equation}
The IPR is usually used to evaluate the degree of localization
of electronic states. Delocalized states are expected to present
small IPR, of order of $N^{-1}$, while localized states have 
larger IPR.  

\section{Results and Discussions}

As typical values of physical parameters we take $V=25\,$meV,
$V'=28\,$meV and $a=10\,$\AA. To simplify the numerical analysis
and to facilitate direct comparison with previous results
in periodic SLs \cite{Sun}, we define two dimensionless quantities. 
We then introduce the reduced energy $\varepsilon = E/V$ and the
reduced electric field $\tilde{F}=eFa/V$. In addition, we focus our 
attention in the first allowed miniband of the SL, which
ranges from $\varepsilon = 0.9$ to $\varepsilon = 14.2$ in the periodic
SL with our chosen parameters.

\begin{figure}

\centerline{\psfig{figure=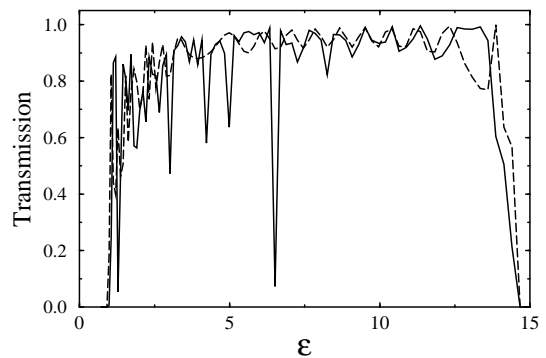,angle=270,width=8cm}} \vspace{6mm}
\caption{Transmission coefficient as a function of the electron energy 
in periodic (dashed line) and Fibonacci superlattices (solid line) under flat 
band conditions. The number of barriers is $N=F_{13}=377$ in both cases.} 
\label{fig1}

\end{figure}

{}Figure~\ref{fig1} collects our results on transmission in both
both periodic and Fibonacci SLs with $N=F_{13}=377$ under flat band
conditions. Results demonstrate
that the miniband structure is changed when quasiperiodicity
is introduced in the SL. Notice the occurrence of gaps within the
allowed miniband, which are absent in the periodic SL. This 
phenomenon is by now well understood (see Ref.~\cite{Macia1}
and references therein), and is due to the reduction of the resonant 
coupling between neighbouring quantum wells. We should also mention 
that this picture does not change very much on increasing the
SL length due to the so-called asymptotic stability \cite{Macia1}, 
i.\ e.\ the global properties can be obtained in practice by considering
very short approximants to the FSL, whereas increasing the
system length leads to change only finer details.

\begin{figure}

\centerline{\psfig{figure=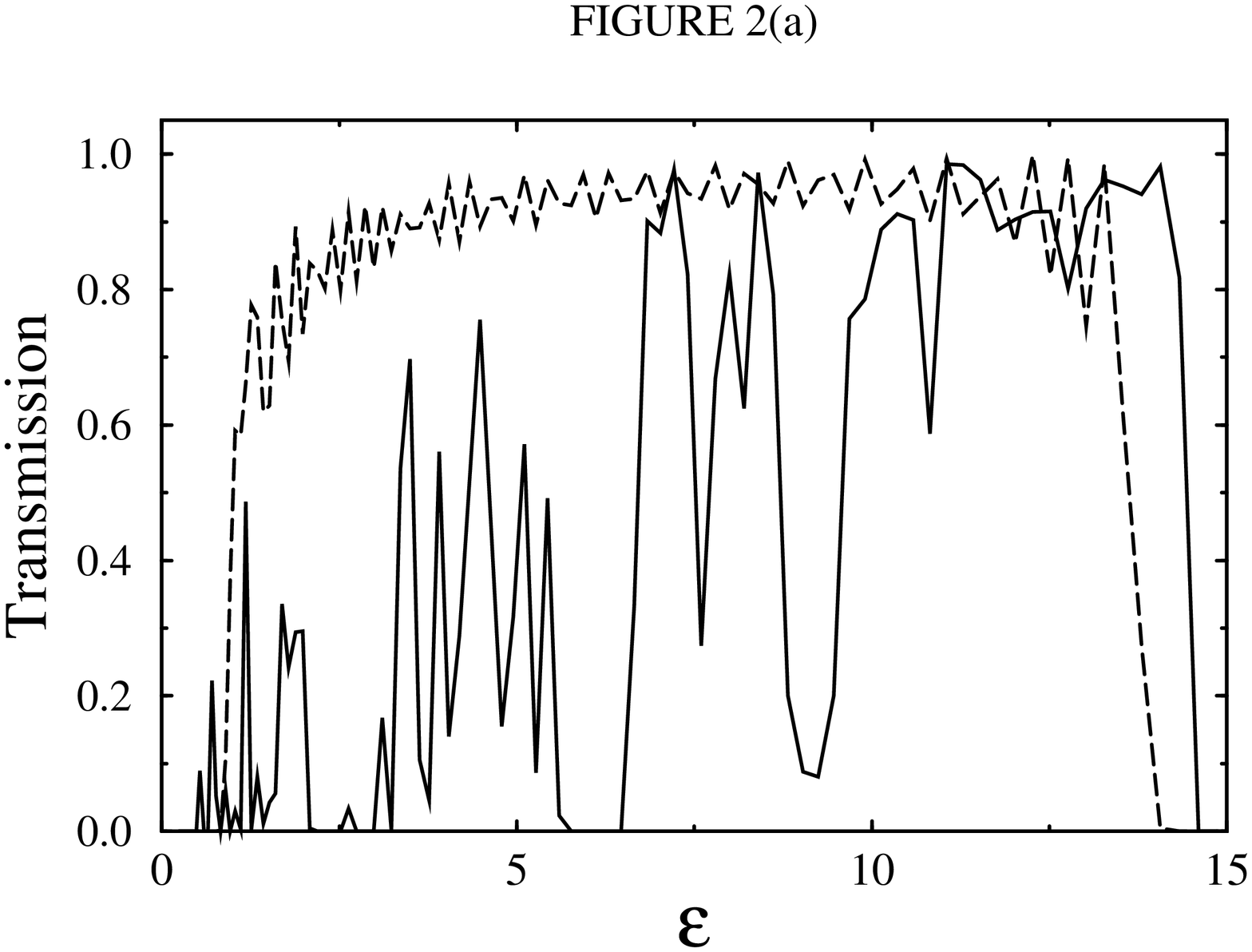,angle=270,width=8cm}} \vspace{6mm}
\centerline{\psfig{figure=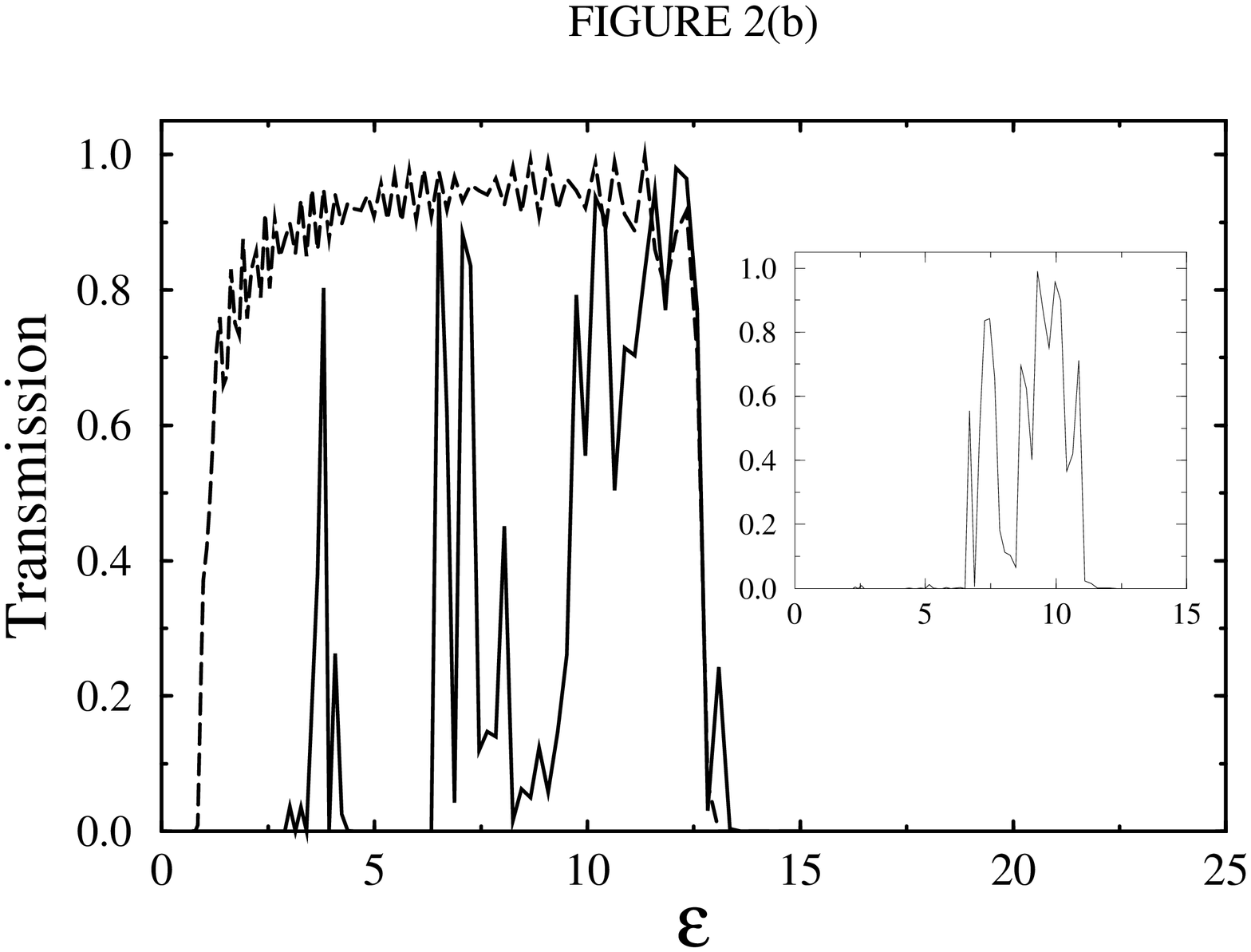,angle=270,width=8cm}} \vspace{6mm}
\caption{Transmission coefficient as a function of the electron energy 
in periodic (dashed lines) and Fibonacci (solid lines) SLs with 
(a) $N=144$ and (b) $N=377$ barriers. The applied electric field is 
$\tilde{F}=0.005$ in all cases. Inset shows the transmission coefficient 
for the FSL with $N=F_{13}=377$ at $\tilde{F}=0.01$} 
\label{fig2}

\end{figure}

The above scenario under flat band conditions is no longer valid
when moderate electric fields are applied. Fig.~\ref{fig2} shows
the results when $\tilde{F}=0.005$ for both periodic and Fibonacci SLs
with different lengths. In the case of periodic SL, only minor 
differences can be detected on increasing the lenght [see 
Fig.~\ref{fig2}(a) and (b)], besides a decrease of the upper
miniband-edge. On the contrary, marked differences appear in 
the case of FSLs when increasing the length when an applied 
electric field is applied. By comparing Fig.~\ref{fig2}(a) and (b), we
observe a strong reduction of transmission properties, especially
at the lower miniband-edge, where several transmission peaks
completely disappear. Thus we are led to the conclusion that
the interplay between the long-range order of FSLs and the 
localization properties of the electric field effects are more
complex than in periodic SLs. These differences clearly arise
from the very different nature of the eigenstates in both types
of SLs \cite{Macia2}. In particular, we deduce from the above 
results that the localization properties of the electric field 
are more pronounced in the case of FSLs. On increasing the electric
field the miniband shrinks, as can be seen in the inset if 
Fig.~\ref{fig2}(b), especially at low energies. Further increase of 
the electric filed leads to the vanishing of the miniband, as occurs in
periodic SLs \cite{Sun}.

\begin{figure}

\centerline{\psfig{figure=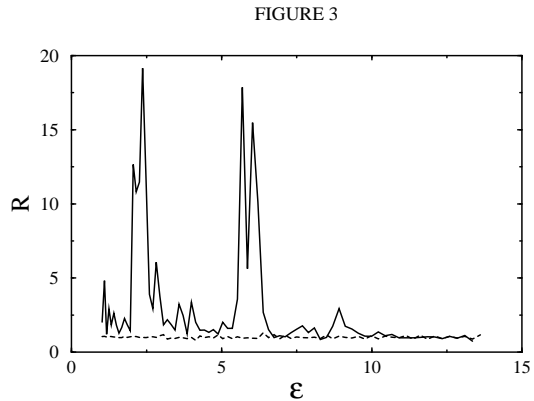,angle=270,width=8cm}} \vspace{6mm}
\caption{Ratio of the IPR in FSL and in periodic SL as a function of the 
electron energy with $N=F_{11}=144$ (dashed line) and $N=F_{13}=377$ 
(solid line) barriers. The applied electric field is $\tilde{F}=0.005$ 
in both cases.} 
\label{fig3}

\end{figure}

To obtain further confirmation of the above statement we have
studied the degree of localization of wave functions. In order
to facilitate comparison between the localization under flat
band and subject to an applied electric field, we define a new
parameter $R$ as the ratio of the IPR of the FSL and the IPR of
the periodic SL at the same values of the energy, electric field
and system length. Results are collected in Fig.~\ref{fig3} for
$N=F_{11}=144$. At flat band the value of the parameter $R$ fluctuates 
around unity, as it can be seen in Fig.~\ref{fig3}. This 
implies that the degree of localization is similar in both
kinds of SL. This agree with previous observations that 
electronic states spread over the entire Fibonacci system,
although the nature of those states is completely different
from those in periodic systems \cite{Macia2}. However, as soon
as the electric field is applied, electrons are much more
localized in the FSL than in the periodic SL, as demonstate
the dramatic increase of the parameter $R$ shown in Fig.~\ref{fig3}.
Notice that the peaks of $R$ correspond to the gaps revealed in the 
trasmission coefficient. This observation reinforces our claim that the 
localization effects of the electric field are stronger in FSL. 

\section{Conclusion}

In this paper we have studied the transmission properties of 
electrons in Fibonacci semiconductor superlattices subject to
an electric field. Results have been compared to those obtained
in periodic superlattices with the same parameters. By taking
the approximation of narrow barriers, which is of interest
when resonant coupling between neigbouring barriers takes 
place and consequently minibands arise, we have been able to 
obtain a closed expression of the trasmission coefficient
within the transfer matrix formalism. In addition, the same 
approach allows us to discuss the degree of localization 
of wave functions by means of the inverse participation ratio.
We have found that at flat band small gaps appear within the 
lower miniband as soon as quasiperiodicity is introduced in the 
sample. However, the Fibonacci SL still presents good transmission
properties, which is consistent with the fact that electronic
states spread over the entire sample in the absence of dc
voltage. On the contrary, dramatic changes occur whenever the
dc field is switched on. In particular, we have found a 
strong reduction of the transmission properties, this reduction
being more significant as the length of the system increases.
The IPR confirms these results and points out that electronic
states are much more localized in Fibonacci SLs than in periodic
ones under the same bias conditions. For large enough electric
fileds one could expect that the fragmented pattern of the
electronic spectrum will be severely modified. We believe that
this result is very important from a practical point of view since
it demonstrates that changes in the electronic structure
should be taken into account when samples are driven by a dc
field.

\acknowledgments 
 
The authors thank Enrique Maci\'{a} and Angel S\'{a}nchez for helpful 
comments.  This work has been supported by CICYT (Spain) under project 
MAT95-0325. 
 

\end{document}